\documentclass[aps, twocolumn, superscriptaddress]{revtex4} 

\usepackage[switch]{lineno}
\usepackage{amsmath}
\usepackage{amssymb}
\usepackage{empheq}
\usepackage[parfill]{parskip}
\usepackage[active]{srcltx}
\usepackage{color}
\usepackage{array}
\usepackage{booktabs}
\usepackage{amsfonts}
\usepackage{dsfont}
\usepackage{graphicx}

\begin{document}

\setlength{\parindent}{0.5cm}

\title{Unsteady phase waves in the 1D swarmalator model with inertia}

\author{Kevin O'Keeffe}
\affiliation{Starling research institute} 

\begin{abstract}
We study a one-dimensional swarmalator model with inertia. Previous studies have focused almost exclusively on the overdamped limit. We find inertia introduces a new unsteady collective state in which the rainbow order parameters undergo multiharmonic oscillations. This ``thrashing'' phase wave bifurcates from the model’s static phase wave state through a subcritical Hopf bifurcation that coincides with a saddle–node of limit cycles. The wave itself exists in clockwise and counterclockwise symmetric pairs. For small populations we observe attractor switching between these chiral states, while for larger systems the dynamics settle onto a single branch.
\end{abstract}

\maketitle
\section{Introduction}
Swarmalators are generalizations of phase oscillators that both move around in space and synchronize the timing of their internal oscillations. They were introduced to model systems in which synchronization and self-assembly interact, such as vinegar eels \cite{quillen2021metachronal}, magnetic domain walls \cite{hrabec2018velocity}, Japanese tree frogs \cite{aihara2014spatio}, colloidal micromotors \cite{yan2012linking,liu2021activity,zhang2020reconfigurable,bechinger2025tunable,heuthe2025tunable}, embryonic cells \cite{tsiairis2016self}, active spheres \cite{riedl2023synchronization} and swarms of robots \cite{barcis2019robots,barcis2020sandsbots,beattie2025realizing,xu2025navigation,quinn2025decentralised}.

The first swarmalator studies considered identical units moving in 2D with mean-field coupling between position and phase \cite{o2017oscillators}. Even this minimal model exhibits a rich variety of collective states, including synchronized clusters, phase waves, and vortex-like structures. Subsequent work has explored the effects of noise, forcing \cite{hong2023thermal,riedl2023synchronization}, time-delayed interactions \cite{blum2024delay,lambu2026delay}, external periodic forcing \cite{anwar2024forcing,anwar2025forcing}, higher-harmonic and higher-order couplings \cite{senthamizhan2024harmonic,anwar2024higherorder}, pulse-coupled interactions \cite{yadav2025pulse}, and higher-dimensional spatial dynamics \cite{yadav2024highdim}. Others have investigated the role of spatial interaction structure, including finite-range coupling, network or topological interactions, and other effects \cite{sar2025range,sar2025control,gou2025topological,keun2025circular,gou2026topological,yu2025swarmalator,horvath2025stigmergic,ghosh2026emergent,yadav2025collective,senthamizhan2025frustration,yu2025collective,lu2025self,o2018ring,lizarraga2024order,kongni2024expected}.

Yet one effect remains unexplored: inertia. Most swarmalator models assume overdamped dynamics for the sake of simplicity. We know from coupled-oscillator theory, however, that inertia can qualitatively change collective behavior. The second-order Kuramoto model, for example, exhibits hysteresis, bistability, and abrupt synchronization transitions \cite{tanaka1997first,acebron2005kuramoto}. How does inertia modify swarmalator dynamics?

In this paper, we address this question using a simple model of identical swarmalators moving on a 1D domain which is especially tractable \cite{o2022collective,yoon2022sync}. We find that inertia fundamentally alters the phase-wave dynamics, giving rise to a new unsteady phase in which the rainbow order parameters undergo mixed–mode oscillations. This ``thrashing'' phase wave emerges from the static phase wave through a subcritical Hopf bifurcation that coincides with a saddle–node of limit cycles. Inertia's main dynamical effect is thus that it generates unsteadiness; the attractors of the non-inertial swarmalator model, recall, are all static \cite{o2022collective}.

\section{Model}
We study the 1D swarmalator model \cite{o2022collective} with inertia terms on both the space and phase dynamics
\begin{align}
m\ddot{x}_i + \dot{x}_i
&=
\omega_i'
+
\frac{J'}{N}\sum_j \sin(x_j-x_i)\cos(\theta_j-\theta_i),
\\
m\ddot{\theta}_i + \dot{\theta}_i
&=
\nu_i'
+
\frac{K'}{N}\sum_j \sin(\theta_j-\theta_i)\cos(x_j-x_i),
\end{align}
where \(x_i\in S^1\) is the position of the \(i\)-th swarmalator on the unit
circle and \(\theta_i\in S^1\) is its phase. The parameters
\((\omega_i',\nu_i')\) are the natural frequencies, $(J',K')$ are coupling
constants, and \(m>0\) is an inertial mass. In sum/difference coordinates $\xi_i := x_i+\theta_i, \eta_i = x_i-\theta_i$ the model becomes 
\begin{align}
m\ddot{\xi}_i+\dot{\xi}_i
&=
\omega_i
+
K r \sin(\phi-\xi_i)
+
J s \sin(\psi-\eta_i),
\\
m\ddot{\eta}_i+\dot{\eta}_i
&=
\nu_i
+
J r \sin(\phi-\xi_i)
+
K s \sin(\psi-\eta_i),
\end{align}
where
\begin{align}
U = r e^{i\phi} := \langle e^{i\xi}\rangle, \qquad
V = s e^{i\psi} := \langle e^{i\eta}\rangle,
\end{align}
and $(\omega,\nu)=(\omega'+\nu',\,\omega'-\nu')$ and $(J,K)=(J'+K')/2,\,(J'-K')/2$.
\begin{figure*}[t!]
\centering
\includegraphics[width = 2 \columnwidth]{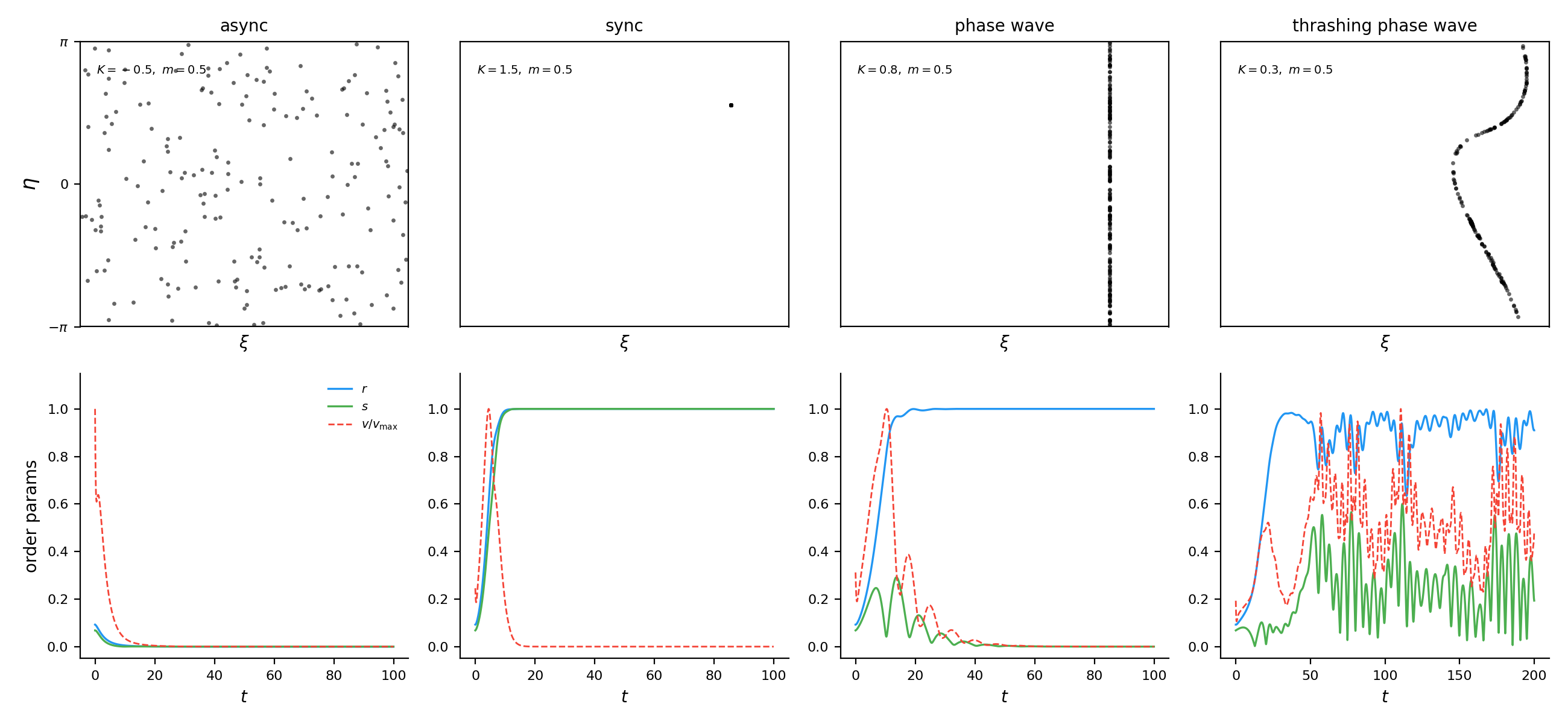}
\caption{Collective states of the inertial 1D swarmalator model, Eqs.~(1)--(2)..
Top row: snapshots of swarmalators in the \((\xi_i,\eta_i)\) plane.
Bottom row: time series of the rainbow order parameters
\(r=|\langle e^{i\xi}\rangle|\), \(s=|\langle e^{i\eta}\rangle|\), and the
normalized mean speed \(v/v_{\max}\).
From left to right: async, sync, phase wave, and thrashing phase wave.
Parameter values \((K,m)\) are indicated in each panel. Simulations used \(N=200\), \(dt=0.05\), and \(m=0.5\).}
\label{states}
\end{figure*}
The quantities \(U\) and \(V\) are the rainbow order parameters, which
generalize the Kuramoto synchronization order parameter
\(Z:=\langle e^{i\theta}\rangle\).

The clean, symmetric structure of the \((\xi,\eta)\) equations makes the 1D
swarmalator model one of the few mobile-oscillator systems amenable to exact
analysis \cite{yoon2022sync,global_sync,o2025stability}. Researchers have  used it to analytically study the effects of disordered coupling \cite{o2022swarmalators,hao2023attractive}, phase frustration \cite{lizarraga2023synchronization}, random pinning \cite{sar2023pinning,sar2024solvable,sar2023swarmalators}, external forcing \cite{anwar2024forced,anwar2025forced}, time delay \cite{okeeffe2026delay}, and other modifications \cite{sar2025effects,hong2023swarmalators,pulsating,ghosh2025dynamics,anwar2024collective,SenthamizhanGopalChandrasekar2025Swarmalators}. \cite{sar2025interplay} reviews the 1D model and explains how it derives from the 2D model.

Here we focus on the case of identical frequencies which can be set
to zero without loss of generality:
\begin{equation}
\omega_i=\nu_i=0.
\end{equation}
This in turn implies \(\phi=\psi=0\) which makes the governing equations especially simple:
\begin{align}
m\ddot{\xi}_i+\dot{\xi}_i
&=
- K r \sin\xi_i - J s \sin\eta_i,
\\
m\ddot{\eta}_i+\dot{\eta}_i
&=
- J r \sin\xi_i - K s \sin\eta_i.
\end{align}

\section{Numerical results}
\begin{figure}[t!]
\centering
\includegraphics[width = \columnwidth]{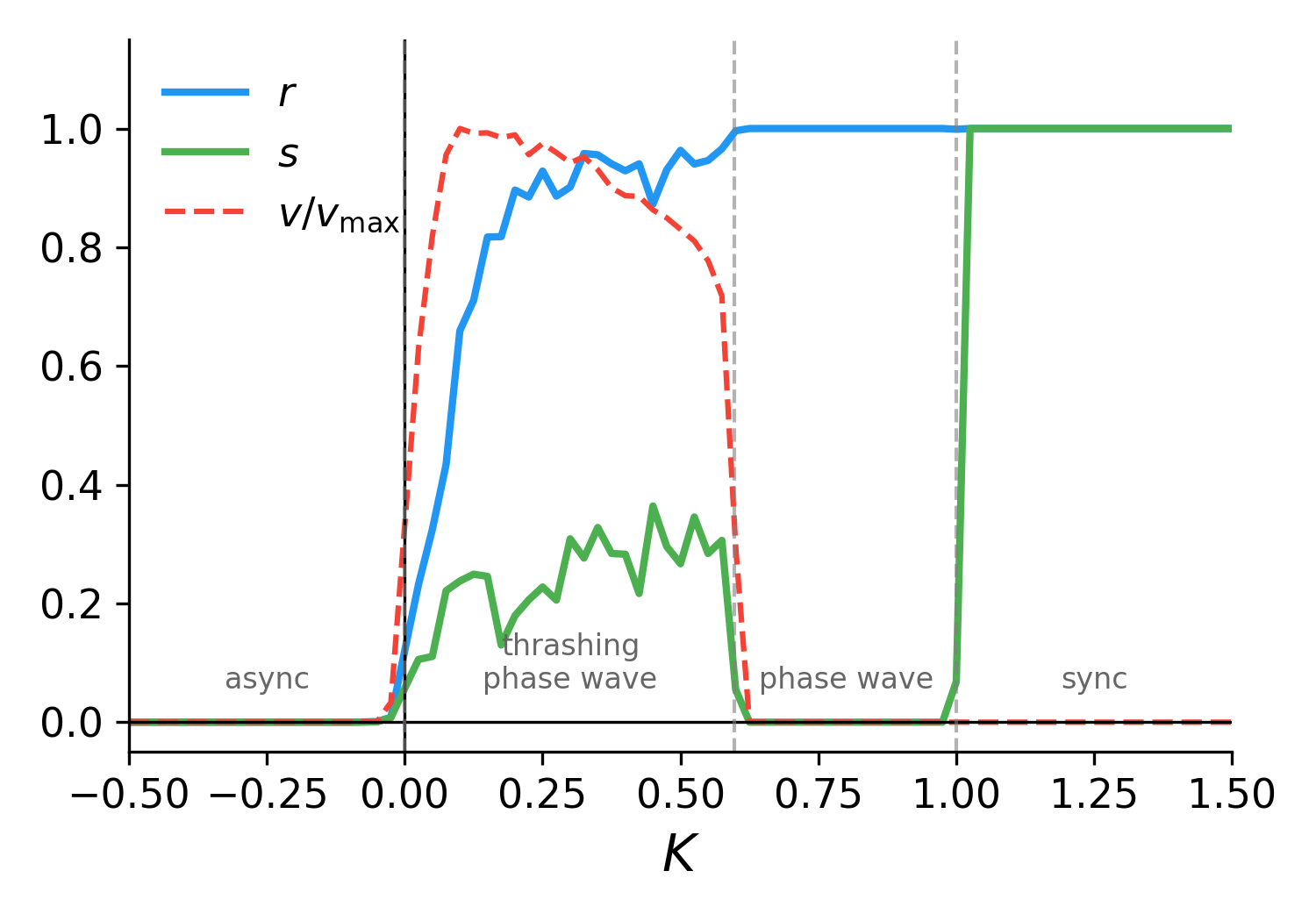}
\caption{Order parameters $r$, $s$, and $v/v_{\max}$ as a function of coupling $K$, for $m=0.5$, $J=1$, $N=500$.  Each point is averaged over 5 random initial conditions. Simulations used RK45 with $dt=0.05$, total time $T=2000$, and transient $T_{\rm trans}=1000$ discarded.}
\label{order_pars}
\end{figure}
By numerically integrating the governing equations, we found that the system
typically settles into one of four collective states. The top row of
Fig.~\ref{states} shows these states as scatter plots in the \((\xi,\eta)\)
plane, while the bottom row shows the corresponding time series of the order
parameters \(r\), \(s\), and the normalized mean speed \(v/v_{\max}\). The four
states are:

\begin{itemize}
    \item \textit{Async}: swarmalators are uniformly distributed in both
    position and phase, so that \((r,s)=(0,0)\). There is no collective order.

    \item \textit{Sync}: all swarmalators share the same position and phase,
    giving \((r,s)=(1,1)\). In the \((\xi,\eta)\) plane this appears as a
    single localized cluster.

    \item \textit{Phase wave}: positions and phases are perfectly correlated,
    so that one of the rainbow order parameters is maximal while the other
    vanishes. For the branch shown here, \((r,s)=(1,0)\), corresponding to a
    \(\xi\)-wave. The symmetry-related \(\eta\)-wave with \((r,s)=(0,1)\) is
    realized equally often from random initial conditions.

    \item \textit{Thrashing phase wave}: a novel oscillatory state in which the
    system remains phase-wave-like on average, but the order parameters undergo
    sustained temporal oscillations. In the example shown, \(r\) remains large
    while \(s\) oscillates periodically about zero. This state has no analogue
    in the overdamped (\(m=0\)) model.
\end{itemize}
Figure~\ref{order_pars} shows the bifurcation structure of the system by plotting \(r\) and
\(s\) versus \(K\) at fixed \(m=0.5\). The normalized mean
velocity \(v/v_{\max}\) is also shown, computed by averaging over both the population and the final $10\%$ of the simulation after transients have died out.

Several trends are clear. For \(K<0\), both order parameters remain near zero,
consistent with the asynchronous state. As \(K\) is increased past zero, the
system enters an oscillatory regime in which the order parameters fluctuate in
time and the mean speed is elevated. Upon further increasing \(K\), the
oscillations disappear and the system settles into a static phase wave, marked
by one rainbow order parameter close to unity and the other close to zero.
Finally, for sufficiently large \(K\), both order parameters approach one,
indicating full synchronization. 

Next we analyze the stability of each state.

\section{Analysis}
\subsection{Sync}
Without loss of generality, we take the synchronized fixed point to be
\begin{equation}
\xi_i^\ast = 0, \qquad \eta_i^\ast = 0,
\end{equation}
for all \(i=1,\dots,N\). At this point,
\begin{equation}
r=s=1.
\end{equation}

To linearize the inertial dynamics, write
\begin{equation}
\xi_i = \delta \xi_i, \qquad \eta_i = \delta \eta_i,
\end{equation}
with \(|\delta \xi_i|, |\delta \eta_i| \ll 1\). Since
\[
U=\langle e^{i\xi}\rangle \approx 1 + i\langle \delta \xi \rangle,
\qquad
V=\langle e^{i\eta}\rangle \approx 1 + i\langle \delta \eta \rangle,
\]
we have, to linear order,
\begin{align}
\mathrm{Im}(Ue^{-i\xi_i}) &\approx \langle \delta \xi \rangle - \delta \xi_i, \\
\mathrm{Im}(Ve^{-i\eta_i}) &\approx \langle \delta \eta \rangle - \delta \eta_i.
\end{align}
Therefore the linearized equations are
\begin{align}
m \,\delta \ddot{\xi}_i + \delta \dot{\xi}_i
&= K\big(\langle \delta \xi \rangle - \delta \xi_i\big)
 + J\big(\langle \delta \eta \rangle - \delta \eta_i\big), \\
m \,\delta \ddot{\eta}_i + \delta \dot{\eta}_i
&= J\big(\langle \delta \xi \rangle - \delta \xi_i\big)
 + K\big(\langle \delta \eta \rangle - \delta \eta_i\big).
\end{align}

It is convenient to separate the mean modes from the shape modes. Define
\begin{equation}
\bar{\xi} := \langle \delta \xi \rangle, \qquad \bar{\eta} := \langle \delta \eta \rangle.
\end{equation}
Averaging the linearized equations over \(i\) gives
\begin{align}
m \ddot{\bar{\xi}} + \dot{\bar{\xi}} &= 0, \\
m \ddot{\bar{\eta}} + \dot{\bar{\eta}} &= 0.
\end{align}
Hence the mean sector contributes the eigenvalues
\begin{align}
\lambda &= 0, \\
\lambda &= -1/m.
\end{align}
The zero eigenvalue is the neutral mode associated with rotational invariance, while \(-1/m\) is a damped center-of-mass velocity mode.

Next define the zero-mean shape perturbations
\begin{equation}
\tilde{\xi}_i := \delta \xi_i - \bar{\xi}, \qquad
\tilde{\eta}_i := \delta \eta_i - \bar{\eta},
\end{equation}
which satisfy \(\langle \tilde{\xi} \rangle = \langle \tilde{\eta} \rangle = 0\). Their equations are
\begin{align}
m \ddot{\tilde{\xi}}_i + \dot{\tilde{\xi}}_i &= -K \tilde{\xi}_i - J \tilde{\eta}_i, \\
m \ddot{\tilde{\eta}}_i + \dot{\tilde{\eta}}_i &= -J \tilde{\xi}_i - K \tilde{\eta}_i.
\end{align}
These decouple under the change of variables
\begin{equation}
p_i := \tilde{\xi}_i + \tilde{\eta}_i, \qquad
q_i := \tilde{\xi}_i - \tilde{\eta}_i,
\end{equation}
giving
\begin{align}
m \ddot{p}_i + \dot{p}_i + (J+K)p_i &= 0, \\
m \ddot{q}_i + \dot{q}_i + (K-J)q_i &= 0.
\end{align}
Thus the remaining eigenvalues are
\begin{align}
\lambda_{p,\pm}
&=
\frac{-1 \pm \sqrt{\,1-4m(J+K)\,}}{2m}, \\
\lambda_{q,\pm}
&=
\frac{-1 \pm \sqrt{\,1+4m(J-K)\,}}{2m}.
\end{align}

For the scaling used in most of our numerics, \(J=1\), this becomes
\begin{align}
\lambda_{p,\pm}
&=
\frac{-1 \pm \sqrt{\,1-4m(1+K)\,}}{2m}, \\
\lambda_{q,\pm}
&=
\frac{-1 \pm \sqrt{\,1+4m(1-K)\,}}{2m}.
\end{align}

We now determine the stability condition. The \(p\)-sector is stable whenever
\begin{equation}
J+K>0,
\end{equation}
which is automatically satisfied in the synchronized regime. The potentially dangerous mode is the \(q\)-sector. This mode is stable iff
\begin{equation}
K-J>0.
\end{equation}
Hence the sync state is linearly stable for
\begin{equation}
\boxed{
K > K_c = J.
}
\end{equation}
In the normalization \(J=1\), this reduces to
\begin{equation}
\boxed{
K > K_c = 1.
}
\end{equation}

\begin{figure}[t!]
\centering
\includegraphics[width = \columnwidth]{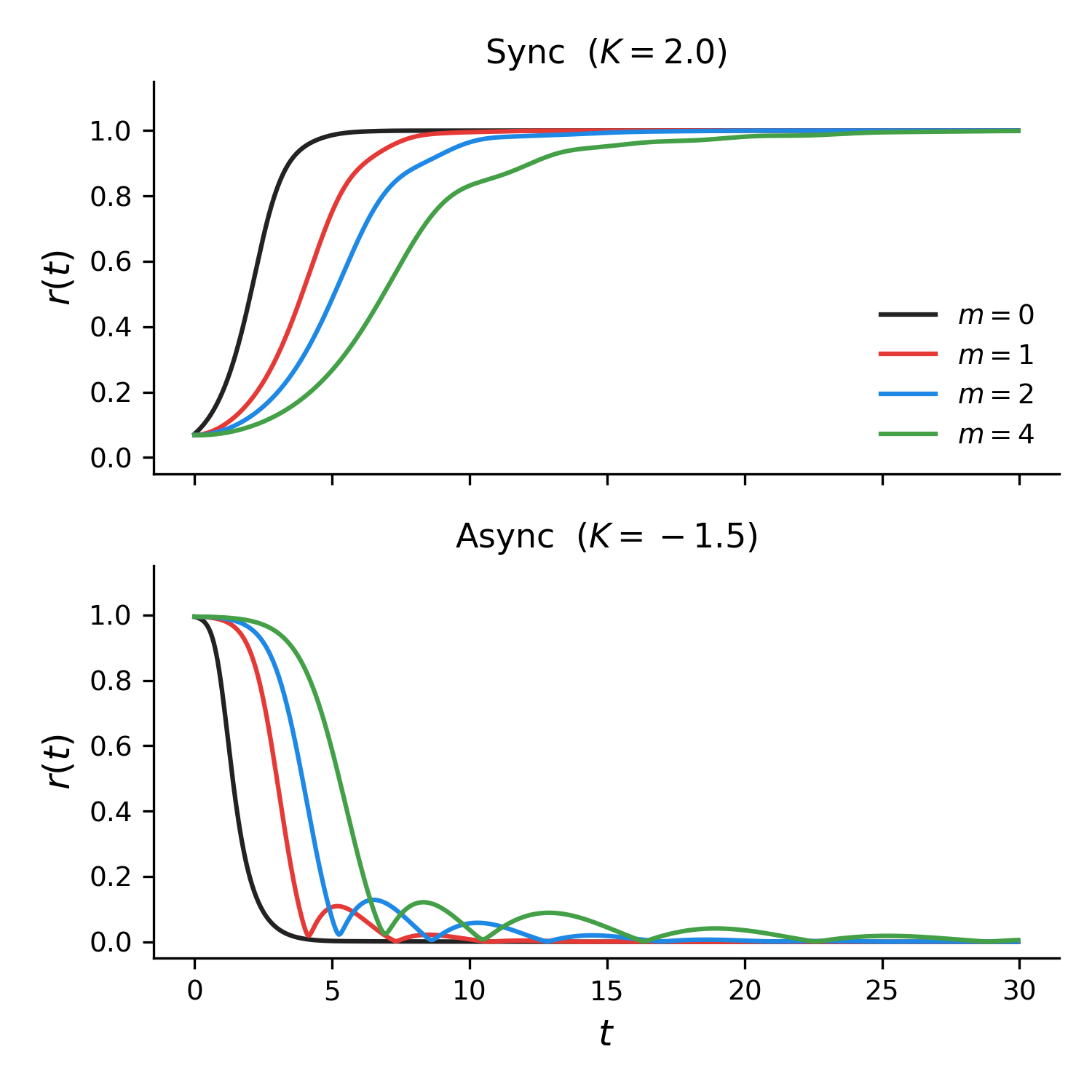}
\caption{Effect of inertia on the transient dynamics of $r(t)$, where we set $r(t)>s(t)$ without loss of generality, starting from random 
initial conditions. Top: sync state ($K=2$); bottom: async state ($K=-1.5$, starting 
near $r=1$). Curves show $m=0,1,2,4,8$. Larger $m$ slows the approach to steady state 
and introduces oscillatory ringing before settling.
Simulations used $N=500$, $dt=0.05$.}
\label{transient}
\end{figure}

Thus, just as in the non-inertial model, the stability threshold of the synchronized state is independent of the inertia \(m\). The mass affects only the transient approach to sync: for sufficiently large \(m\), the relaxation becomes oscillatory (Figure~\ref{transient} top row), but the location of the transition remains unchanged.

\subsection{Async}
Consider the phase-space density
\[
f(\xi,\eta,u,v,t),
\]
which evolves according to the continuity equation
\begin{equation}
\partial_t f + u\partial_\xi f + v\partial_\eta f
+ \partial_u(F_\xi f) + \partial_v(F_\eta f) = 0,
\end{equation}
where
\begin{align}
F_\xi &= \frac{-u + K\,\mathrm{Im}(U e^{-i\xi}) + J\,\mathrm{Im}(V e^{-i\eta})}{m}, \\
F_\eta &= \frac{-v + J\,\mathrm{Im}(U e^{-i\xi}) + K\,\mathrm{Im}(V e^{-i\eta})}{m}.
\end{align}
The order parameters are
\[
U=\iint e^{i\xi} f\,d\xi d\eta dudv,
\qquad
V=\iint e^{i\eta} f\,d\xi d\eta dudv.
\]

The asynchronous state corresponds to a distribution uniform in
\((\xi,\eta)\) with zero velocities,
\begin{equation}
f_0(\xi,\eta,u,v)=\frac{1}{(2\pi)^2}\delta(u)\delta(v).
\end{equation}

We perturb this state as
\[
f = f_0 + \epsilon f_1,
\qquad
0<\epsilon\ll1,
\]
and assume normal-mode perturbations of the form
\[
f_1 \propto e^{\lambda t+i\xi}.
\]
This mode corresponds to the first Fourier harmonic that generates the
order parameter \(U\).

Linearizing the continuity equation and integrating over velocities yields
\begin{equation}
m\lambda^2 a_{10} + \lambda a_{10} = \frac{K}{2} a_{10},
\end{equation}
where \(a_{10}\) is the amplitude of the \((1,0)\) Fourier mode of the spatial density.
An identical equation is obtained for the \((0,1)\) mode associated with \(V\).
Importantly, the cross-coupling \(J\) does not appear because the
corresponding terms vanish when projected onto the first Fourier harmonics.

Seeking solutions \(a_{10}\propto e^{\lambda t}\) gives the characteristic equation
\begin{equation}
m\lambda^2+\lambda-\frac{K}{2}=0,
\end{equation}
with eigenvalues
\begin{equation}
\lambda_\pm =
\frac{-1 \pm \sqrt{1+2mK}}{2m}.
\end{equation}

The stability threshold follows immediately. If \(K>0\), then
\(\lambda_+>0\) and the asynchronous state is unstable.
If \(K<0\), both eigenvalues have negative real part and the
state is linearly stable. Hence the critical coupling is
\begin{equation}
\boxed{K_c=0.}
\end{equation}

Thus the stability threshold of the asynchronous state is unchanged by
inertia and is independent of the cross-coupling \(J\).
Inertia affects only the transient dynamics (Figure~\ref{transient} bottom row).

\subsection{Phase wave}

Without loss of generality, we consider the $\xi$–wave, for which
\begin{equation}
\xi_i^\ast = 0, \qquad 
\eta_i^\ast = \frac{2\pi i}{N},
\qquad i=1,\dots,N .
\end{equation}
In this state,
\begin{equation}
r=1, \qquad s=0,
\end{equation}
so the swarmalators are perfectly ordered in the $\xi$ direction and uniformly
distributed in $\eta$.

We perturb the phase wave as
\begin{equation}
\xi_i = \delta \xi_i, \qquad 
\eta_i = \eta_i^\ast + \delta \eta_i,
\end{equation}
with $|\delta\xi_i|,|\delta\eta_i|\ll1$. Expanding the order parameter
\[
U = \frac{1}{N}\sum_{j=1}^N e^{i\xi_j}
\approx 1 + i\langle\delta\xi\rangle,
\]
gives
\begin{equation}
\mathrm{Im}(Ue^{-i\xi_i})
\approx
\langle\delta\xi\rangle - \delta\xi_i .
\end{equation}

Because the base state is uniform in $\eta$, the instability is governed by the
first Fourier harmonic of the $\eta$ distribution. We therefore write
\begin{equation}
\delta\xi_i = a\cos\eta_i^\ast + b\sin\eta_i^\ast, \qquad
\delta\eta_i = c\cos\eta_i^\ast + d\sin\eta_i^\ast .
\end{equation}

Substituting these expressions into the linearized equations and retaining only
first-harmonic terms yields two identical $2\times2$ systems. For the cosine
sector we obtain
\begin{align}
m\ddot a + \dot a &= -K a + \frac{J}{2}c, \\
m\ddot c + \dot c &= -J a + \frac{K}{2}c ,
\end{align}
while the sine sector obeys the same equations with $(a,c)$ replaced by $(b,d)$.

Seeking normal modes $a,c\propto e^{\lambda t}$ gives
\begin{equation}
\begin{pmatrix}
m\lambda^2+\lambda+K & -J/2 \\
J & m\lambda^2+\lambda-K/2
\end{pmatrix}
\begin{pmatrix}
a \\ c
\end{pmatrix}
=0 .
\end{equation}
Nontrivial solutions require
\begin{equation}
\left(m\lambda^2+\lambda+K\right)
\left(m\lambda^2+\lambda-\frac{K}{2}\right)
+\frac{J^2}{2}=0 .
\end{equation}

Defining
\begin{equation}
x := m\lambda^2+\lambda ,
\end{equation}
this can be written compactly as
\begin{equation}
2x^2 + Kx + (J^2-K^2)=0 .
\label{eq:pw-char}
\end{equation}
\begin{figure}[t!]
\centering
\includegraphics[width = 0.8 \columnwidth]{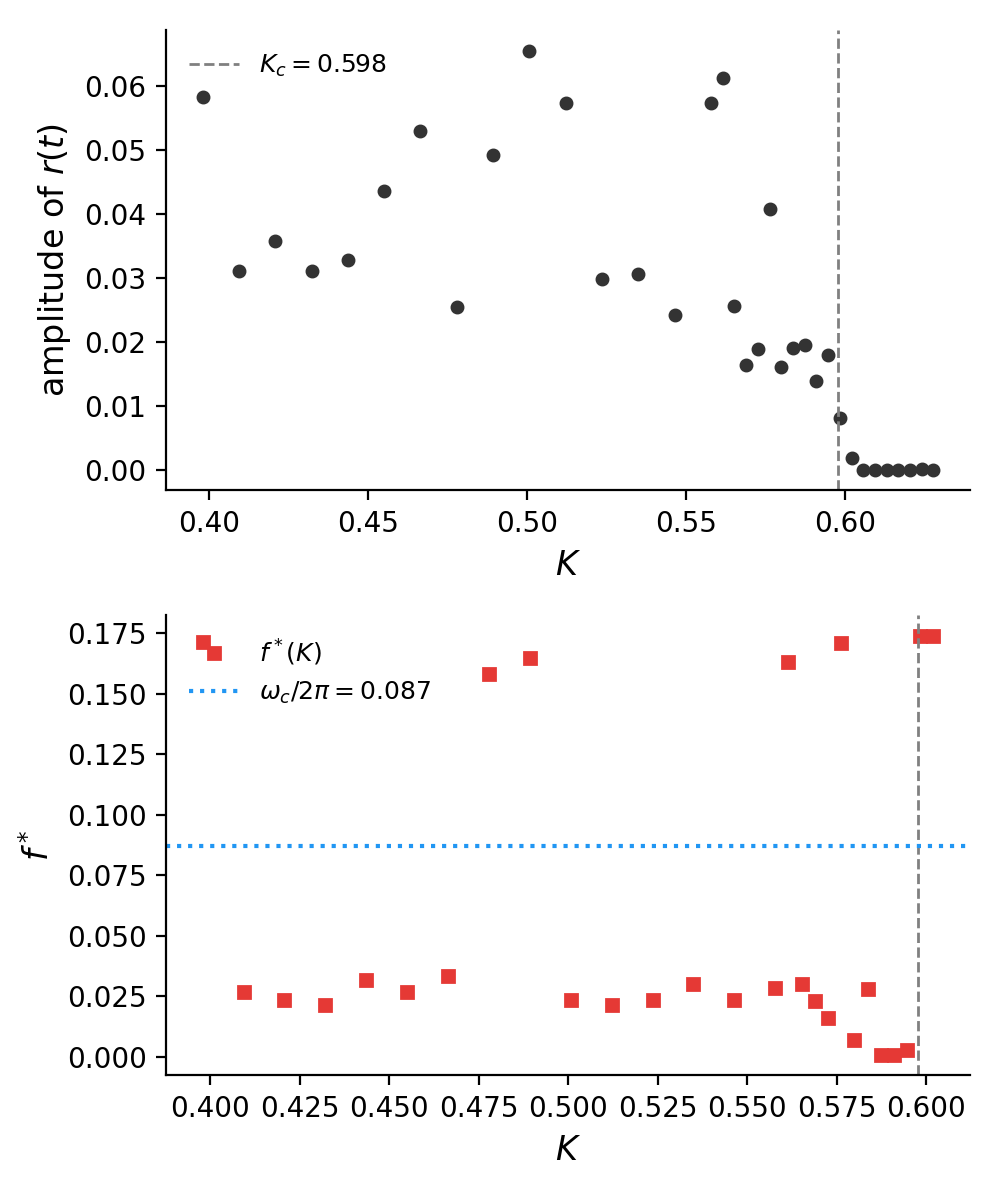}
\caption{Evidence for subcritical Hopf bifurcation at $m=0.5$, $N=100$.
(a) Oscillation amplitude of $s(t)$ vs $K$ from random initial conditions,
showing a finite jump at $K_c \approx 0.598$ rather than the continuous
$\sqrt{K_H - K}$ scaling expected for a supercritical Hopf.
(b) Frequency of oscillation vs $K$, showing a discontinuous jump at $K_H$.}
\label{subcritical}
\end{figure}

To determine when the phase wave loses stability we look for a Hopf
bifurcation by setting
\begin{equation}
\lambda=i\omega, \qquad \omega>0 .
\end{equation}
Then
\[
x=-m\omega^2+i\omega .
\]
Substituting into Eq.~\eqref{eq:pw-char} and separating real and imaginary
parts gives
\begin{align}
-4m\omega^3 + K\omega &= 0, \\
2\left(m^2\omega^4-\omega^2\right)-Km\omega^2+(J^2-K^2) &= 0 .
\end{align}
From the imaginary part the nontrivial Hopf frequency is
\begin{equation}
\omega_c^2 = \frac{K}{4m}.
\end{equation}
Substituting this into the real part yields the Hopf boundary
\begin{equation}
\boxed{
m_H(K) = \frac{4K}{8J^2-9K^2}.
}
\end{equation}

Equivalently, for fixed $m$, the critical coupling satisfies
\begin{equation}
9mK_c^2 + 4K_c - 8mJ^2 = 0 ,
\end{equation}
so that the positive branch is
\begin{equation}
\boxed{
K_H(m) =
\frac{2\!\left(\sqrt{1+18m^2J^2}-1\right)}{9m}.
}
\end{equation}
In the normalization $J=1$ this becomes
\begin{equation}
K_H(m) =
\frac{2\!\left(\sqrt{1+18m^2}-1\right)}{9m}.
\end{equation}
This prediction agrees well with numerics. For example, at $m=0.5$ and $J=1$
the formula gives
\[
K_H \approx 0.598 ,
\]
matching the transition observed in Fig.~\ref{order_pars}.
\begin{figure}[t!]
\centering
\includegraphics[width = 0.8 \columnwidth]{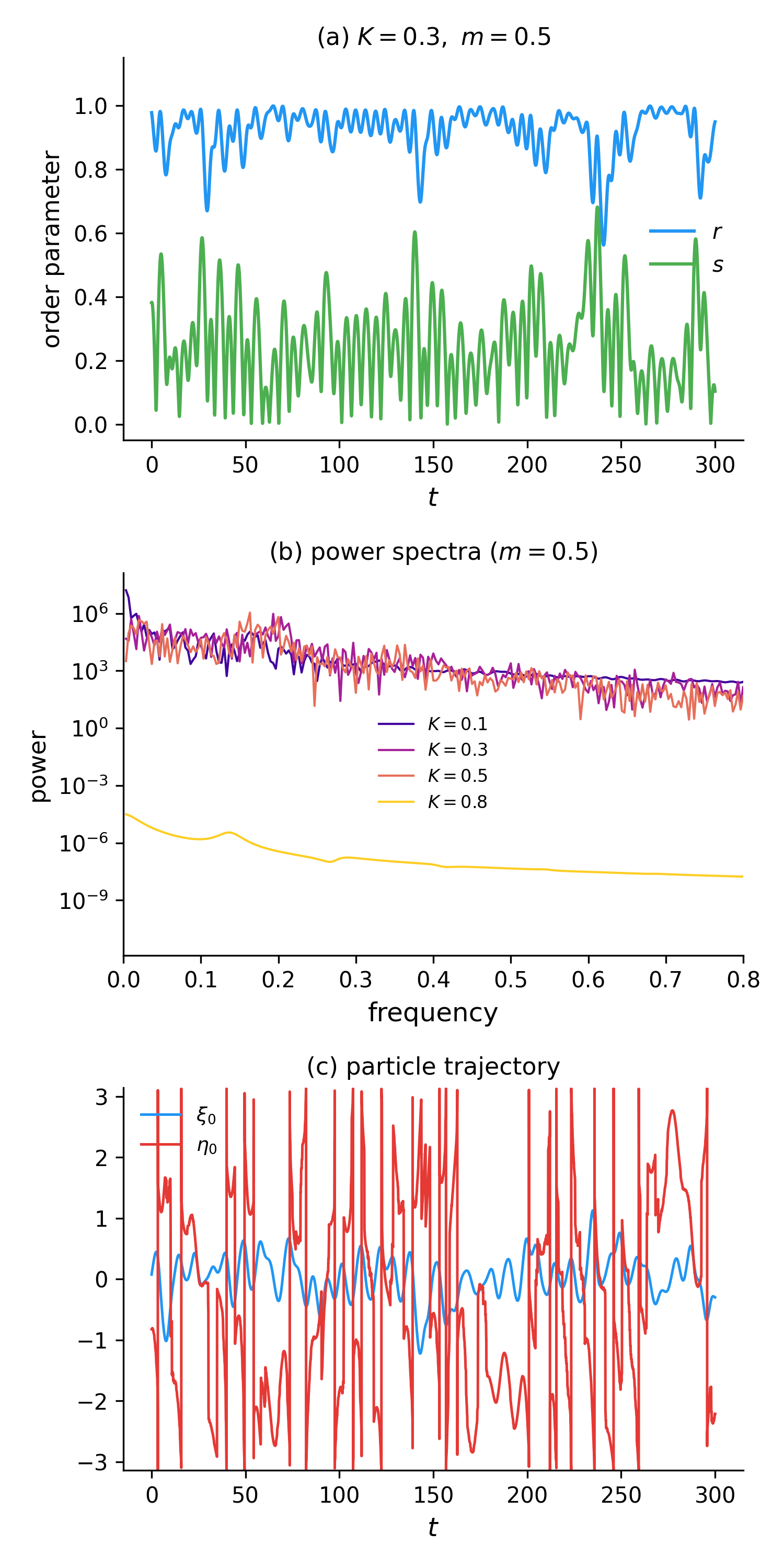}
\caption{Characterization of the thrashing phase wave at $K=0.3$, $m=0.5$.
(a) Time series of $r(t)$ and $s(t)$.
(b) Power spectrum of $s(t)$, with dominant frequency $f^* \approx 0.20$.
(c) Trajectories $\xi_i(t)$ (solid) and $\eta_i(t)$ (dashed) for a typical swarmalator.
Simulations used $N=500$, $dt=0.05$, with the first $100$ time units discarded as transient.}
\label{breathing}
\end{figure}
\begin{figure}[t!]
\centering
\includegraphics[width = 0.9 \columnwidth]{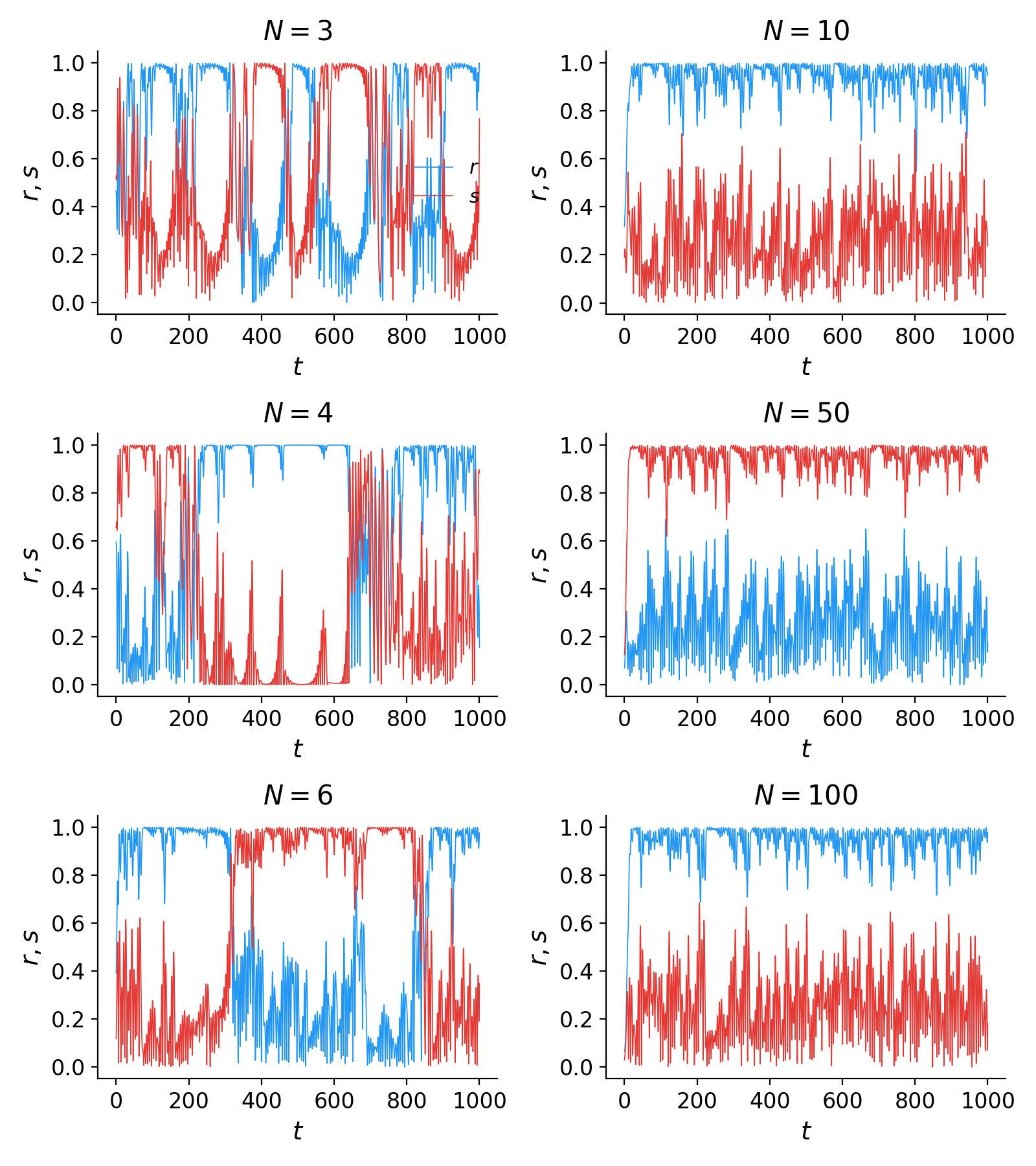}
\caption{Attractor switching between the clockwise ($r>s$) and counter-clockwise ($r<s$) phase wave at $K=K_c - 0.05$, $m=0.5$, $J=1$. For small $N$, left column, the system switches intermittently between the two states. For large $N$, right column, the switching disappears and the system settles into a single phase wave.}
\label{switching}
\end{figure}

The Hopf bifurcation is subcritical. We confirmed this by computing the first
Lyapunov coefficient $l_1$ numerically for $N=3,\dots,10$ and in all cases
found $l_1>0$. Moreover, when sweeping $K$ through the Hopf point we observe
discontinuous jumps in both oscillation amplitude and frequency
(Fig.~\ref{subcritical}). This rules out a supercritical Hopf, for which the
oscillation amplitude would grow continuously from zero, and is consistent
with a nearby saddle--node of limit cycles. Numerically we find that the
critical parameter value $K_{SN}$ appears to coincide with the Hopf point
$K_{SN}\approx K_H$. We also scanned for bistability between the static and
thrashing phase waves but found none.

\subsection{Thrashing phase wave}
This state arises for $0 < K \lesssim K_H(m)$. The order parameters $r$ and $s$ oscillate in time rather than 
settling to constant values. Figure~\ref{breathing} characterizes 
the state at $K=0.3$, $m=0.5$. Panel (a) shows the time series 
$r(t)$ and $s(t)$: both order parameter undergoes sustained oscillations. Panel (b) shows the power 
spectrum of $s(t)$, which exhibits a sharp peak at dominant 
frequency $f^* \approx 0.20$ (period $\approx 5.1$), indicating a 
well-defined collective oscillation. Panel (c) shows representative 
particle trajectories $\xi_i(t)$ and $\eta_i(t)$, illustrating how 
the swarm slowly drifts while the relative phase relations oscillate, 
producing the macroscopic breathing of the order parameters.

For small system sizes we observe an additional phenomenon: 
switching between symmetry-related phase-wave states. The left 
column of Fig.~\ref{switching} shows that for small $N$ the system 
alternates between clockwise and counterclockwise phase waves, 
corresponding to intervals with $r>s$ and $s>r$, respectively. The switching becomes increasingly rare as $N$ increases and appears to vanish in the large-$N$ limit, where the system remains 
locked into a single chirality. 

The effect is robust to the choice of numerical integrator. 
Figure~\ref{robust} in the Appendix shows consistent results across 
three high-order adaptive schemes: RK45 (Dormand–Prince, 4/5 order), 
DOP853 (Dormand–Prince, 8th order), and Radau (implicit Runge–Kutta, 
5th order), confirming that the switching is a genuine dynamical 
feature, and not a numerical artifact.

\begin{figure}[t!]
\centering
\includegraphics[width = \columnwidth]{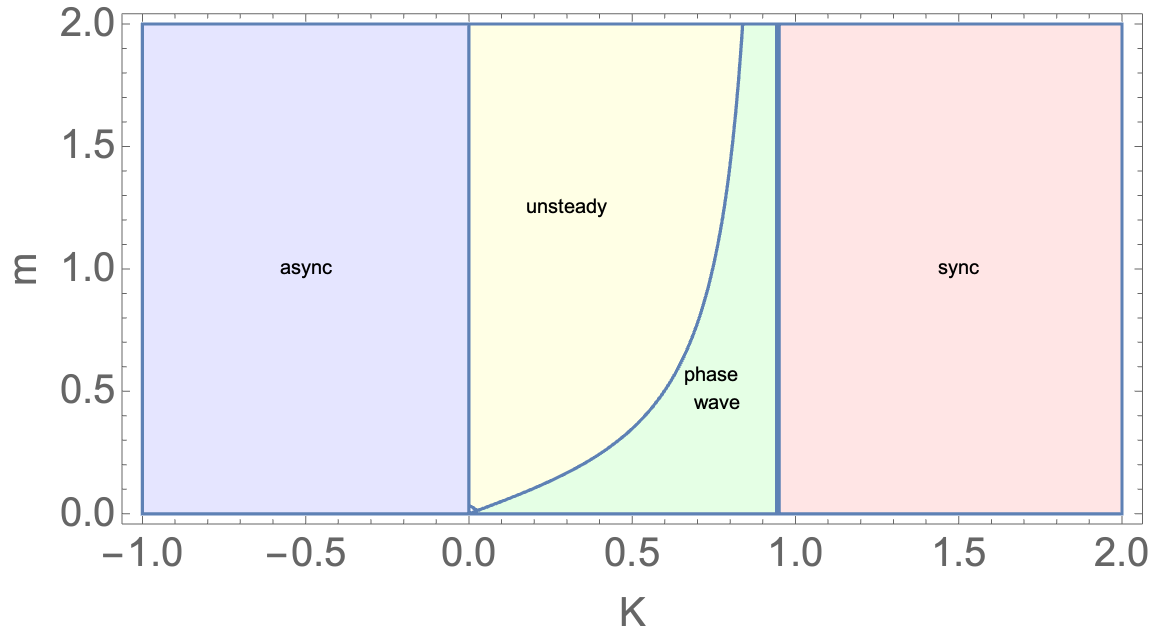}
\caption{State diagram in $(K, m)$ space. The thick black curves correspond to analytic predictions.}
\label{bif-diagram}
\end{figure}

Figure~\ref{bif-diagram} summarizes our analytic findings with a bifurcation diagram in $(K,m)$ space. 

\section{Discussion}
We studied the 1D swarmalator model with inertia. The main
result is simple: inertia leaves the async and sync stability thresholds untouched, but
destabilizes the phase wave. A thrashing phase wave is born, which has multiharmonic oscillations and attractor switching. These findings may be relevant for scientist or engineers exploiting swarmalators for applications \cite{o2019review}.

What's interesting is that inertia-induced unsteadiness does \textit{not} occur in regular oscillators with identical natural frequencies. There, only the static sync and async are realized \cite{tanaka1997first,acebron2005kuramoto} (we emphasive this is when the oscillators are \textit{identical}). Here, the interplay between the sync and self-assembly allows for richer dynamical coupling to the inertial effects.

Future work could add more realism into the model, such as distributed natural frequencies, random couplings, delayed communications, or thermal noise. Exploring the model in two or three spatial dimensions \cite{o2024solvable} may also be fruitful.

\bibliographystyle{apsrev}

\appendix
\section{Robustness of attractor switching}
\begin{figure*}[t!]
\centering
\includegraphics[width = 2\columnwidth]{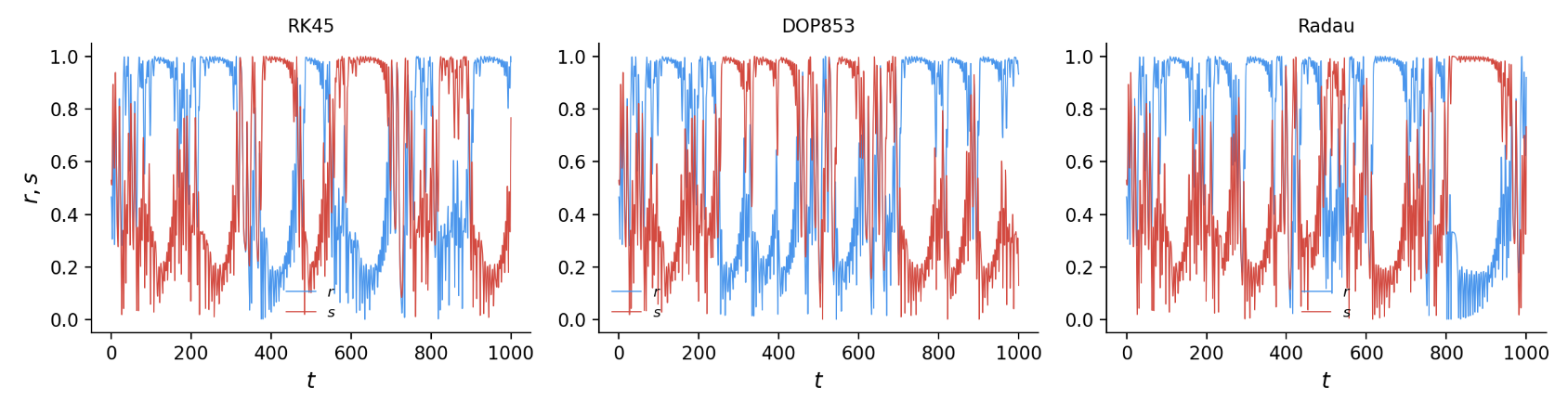}
\caption{Attractor switching for $N=3$, $K=K_H - 0.05$ computed with three independent integrators: RK45, DOP853, and Radau. The switching between the clockwise ($r>s$) and counter-clockwise ($r<s$) phase wave persists across all three, confirming it is not a numerical artifact.}
\label{robust}
\end{figure*}


\end{document}